\documentclass[journal=jacsat,manuscript=article]{achemso}

\usepackage{chemformula} 
\usepackage[T1]{fontenc} 

\usepackage{amsmath}
\usepackage{textcomp}
\usepackage{bm}

\newcommand{\Ea}{\ensuremath{{\cal E}_1}}
\newcommand{\Eb}{\ensuremath{{\cal E}_2}}
\newcommand{\Ec}{\ensuremath{{\cal E}_3}}

\newcommand{\Er}{\ensuremath{{\cal E}_{\rm R}}}
\newcommand{\Eo}{\ensuremath{{\cal E}_{1,2,3}}}

\author{Jacek~Kasprzak}
\affiliation{Universit{\'e} Grenoble Alpes, CNRS, Grenoble INP, Institut N\'{e}el, 38000 Grenoble, France}
\email{jacek.kasprzak@neel.cnrs.fr} 

\author{Daniel~Wigger}
\affiliation{Department of Theoretical Physics, Wroc\l{}aw University of Science and Technology, 50-370~Wroc\l{}aw, Poland}

\author{Thilo~Hahn}
\affiliation{Institut of Solid State Theory, University of M\"unster, 48149 M\"unster, Germany}

\author{Tomasz~Jakubczyk}
\affiliation{Universit{\'e} Grenoble Alpes, CNRS, Grenoble INP, Institut N\'{e}el, 38000 Grenoble, France}
\alsoaffiliation{Institute of Experimental Physics, Faculty of Physics, University of Warsaw,\\ 02-093 Warszawa, Poland}

\author{{\L}ukasz~Zinkiewicz}
\affiliation{Institute of Experimental Physics, Faculty of Physics, University of Warsaw,\\ 02-093 Warszawa, Poland}

\author{Pawe\l~Machnikowski}
\affiliation{Department of Theoretical Physics, Wroc\l{}aw University of Science and Technology, 50-370~Wroc\l{}aw, Poland}

\author{Tilmann~Kuhn}
\affiliation{Institut of Solid State Theory, University of M\"unster, 48149 M\"unster, Germany}

\author{Jean-Fran\c{c}ois~Motte}
\affiliation{Universit{\'e} Grenoble Alpes, CNRS, Grenoble INP, Institut N\'{e}el, 38000 Grenoble, France}

\author{Wojciech~Pacuski}
\affiliation{Institute of Experimental Physics, Faculty of Physics, University of Warsaw,\\ 02-093 Warszawa, Poland}

\title[Mn-spin four wave mixing]
  {Coherent dynamics of a single Mn-doped quantum dot revealed by four-wave mixing spectroscopy}

\abbreviations{QD,FWM,Mn,MBE,PID,FSS,DBR,SIL,NA,PL,FWHM}

\begin{document}




\begin{abstract}
For future quantum technologies the combination of a long quantum state lifetime and an efficient interface with external optical excitation are required. In solids, the former is for example achieved by individual spins, while the latter is found in semiconducting artificial atoms combined with modern photonic structures. One possible combination of the two aspects is reached by doping a single quantum dot, providing a strong excitonic dipole, with a magnetic ion, that incorporates a characteristic spin texture. Here, we perform four-wave mixing spectroscopy to study the system's quantum coherence properties. We characterize the optical properties of the undoped CdTe quantum dot and find a strong photon echo formation which demonstrates a significant inhomogeneous spectral broadening. Incorporating the Mn$^{2+}$ ion introduces its spin-5/2 texture to the optical spectra via the exchange interaction, manifesting as six individual spectral lines in the coherent response. The random flips of the Mn-spin result in a special type of spectral wandering between the six transition energies, which is fundamentally different from the quasi-continuous spectral wandering that results in the Gaussian inhomogeneous broadening. Here, the discrete spin-ensemble manifests in additional dephasing and oscillation dynamics.
\end{abstract}

\section{Introduction}
The incorporation of individual quantum systems into solid-state platforms,\cite{MichlerScience00, WallraffNature04, KoppensNature06, PlaNature12, BrannyNatCom17, EvansScience18} their coherent control, and interfacing them with external degrees of freedom\cite{PettaScience05, NowackScience07, LodahlRMP15, WiggerQUTE21} is a key for implementation of quantum technologies. One of such promising platforms are semiconductor quantum dots (QDs),\cite{MichlerBook03} which owing to constant progress in the epitaxial growth\cite{HermelinNature11, KuhlmannNatCom15} and chemical synthesis,\cite{GarciaScience21} have now reached a tremendous structural quality.\cite{SomaschiNatPhot16} In parallel, processing of this material has been driven virtually to perfection permitting advanced engineering of the light-matter coupling with photonic structures.\cite{LodahlRMP15} As a result, QDs in photonic micro-structures serve as compact sources of single photons for quantum cryptography.\cite{SchimpfSciAdv21}

Conversely, optical or electrical control of single quantum states confined to a QD is challenging, nonetheless intensely pursued in fundamental research.\cite{StinaffScience06, ReiterJoPCM14, KaldeweyPRB17, WeissOptica21} Over the last decade, a major progress has been achieved in measuring\cite{LangbeinPRL05, PattonPRB06, KasprzakNJP13} and controlling\cite{FrasNatPhot16, WiggerOptica18, HenzlerPRL21} the coherence of optical transitions attributed to the bound electron-hole pair, forming a QD exciton. This was achieved by performing coherent ultrafast nonlinear spectroscopy,\cite{LangbeinOL06} in particular four-wave mixing (FWM) on photonic devices hosting InGaAs QDs. However, the exciton radiative lifetime $T_1$ lies typically in the nanosecond range, thus setting the upper bound for its coherence time $T_2\leq 2T_1$. Although an exciton represents an efficient interface between light and matter, its short $T_2$ limits its usage as a qubit. Hence, a promising perspective in this field is the search for efficient coupling schemes between an exciton and quantum systems exhibiting significantly longer $T_2$, for example dark exciton states\cite{PoemNPhy10, KorkusinskiPRB13} or individual spins.\cite{HansonNature08, YilmazPRL10, QuinteiroPRB14, HinzPRB18} Besides employing QDs charged by a single electron\cite{AtatureScience06} or hole\cite{GerardotNature08}, the latter can be achieved by doping QDs with single magnetic ions, like manganese (Mn), which is part of the emerging research area of solotronics, i.e., the field of optoelectronics associated with single dopants.\cite{KoenraadNatMat11, KobakNatComm14}

To bring the benefits of quantum optics and related tools to solotronics, one first needs to introduce the dopant ion into the  QD\cite{BesombesPRL04, KudelskiPRL07, FainblatNanoLet16} and enclose it within a photonic structure to enhance the light-matter coupling.\cite{PacuskiCGD14, PacuskiCGD17} Recently, this requirement was fulfilled by molecular beam epitaxy (MBE) of the II-VI semiconductor CdTe, nowadays offering QD systems hosting various magnetic ions\cite{KobakNatComm14, SmolenskiNatComm16, LafuentePRB16} and reliable fabrication of optical microcavities.\cite{DangPRL98} Nevertheless, the progress in coherent spectroscopy of single excitons in CdTe QDs has been laborious,\cite{PattonPRB06, PacuskiCGD17} due to the restricted availability of femtosecond laser sources emitting in the visible range.

In the present work, we perform FWM spectroscopy of single exciton in CdTe QDs embedded in a microcavity. We first employ a non-magnetic dot, to demonstrate its quantum character by performing the Rabi rotation measurements.\cite{PattonPRL05, WiggerPRB17} Next, we determine the exciton's population and coherence dynamics. In the latter case, we reveal the formation of a photon echo,\cite{LangbeinPRL05} phonon-induced dephasing (PID),\cite{JakubczykACSPhot16, WiggerOL20} and coherence beating owing to the fine-structure splitting (FSS) of the exciton.\cite{KasprzakPRB08, MermillodPRL16} Finally, we focus on a QD doped with an individual Mn$^{2+}$ ion, which in the II-VI material CdTe acts as an isoelectronic impurity. We show that the exciton-Mn$^{2+}$ exchange interaction introduces an additional ensemble characteristic in the time-averaged experiments. In this specific situation, the impact of the Mn$^{2+}$-spin with total spin quantum number $S=5/2$ results in the appearance of six different transition energies associated with the six possible orientations of the Mn spin. It has been shown that due to this characteristic spectral features the Mn spin can be initialized, read out and controlled\cite{LePRB10, BesombesNanophot15} and protocols have been suggested for a selective switching of the spin state.\cite{ReiterPRL09, ReiterPRB11, ReiterPRB12} This report is thus an initial step on the spectroscopic quest towards fully fledged coherent quantum control of possible spin-photon interfaces.\cite{HansonNature08}

\section{Sample and Experiment}
\begin{figure}[h]
    \centering
    \includegraphics[width=0.6\columnwidth]{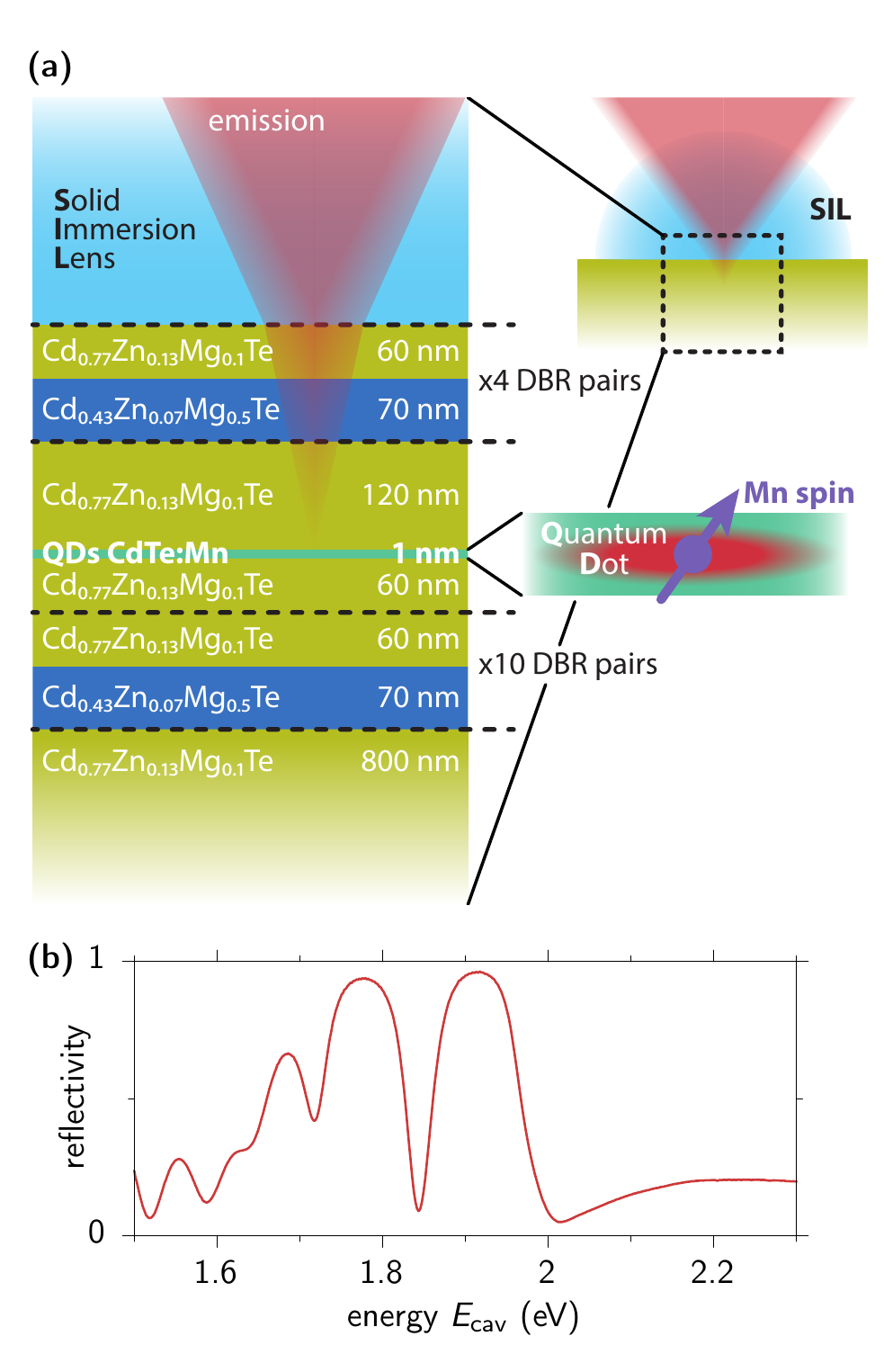}
   \caption{Microcavity sample. (a) Schematic picture of the sample structure including top and bottom distributed Bragg reflectors (DBRs), the Mn-doped quantum dot (QD) layer (highlighted on the right), and a solid immersion lens (SIL) to improve focusing of the laser light and the collection efficiency. (b) Reflectivity spectrum of the microcavity.}
    \label{fig:sample}
\end{figure}%
To perform FWM experiments of a single Mn-doped QD, we specifically conceive the microcavity sample schematically depicted in Fig.~\ref{fig:sample}(a). We have previously shown that in a standard microcavity, the light-matter interaction is enhanced through the intra-cavity field amplification,\cite{FrasNatPhot16} whilst preserving spectral matching with the excitation via femtosecond laser pulses. The asymmetric cavity design permits to reflect almost the entire optical response toward the detection path. With this methodology, we increase the FWM collection efficiency by several orders of magnitude with respect to planar samples. Inspired by that performance-boost, we here go beyond the previously used half-cavity design.\cite{PacuskiCGD17} We develop full monolithic cavities, choosing the quaternary alloy CdZnMgTe as building material. After having deposited a buffer on the GaAs substrate, a bottom distributed Bragg reflector (DBR) is grown by alternating Mg content between 10\% and 50\%. After the completion of 10 layer-pairs for the bottom DBR, we proceed by the formation of the $\lambda$ cavity. At the calculated field antinode we flush a CdTe QD layer and nominally set the Mn concentration to 0.1\% to allow a diluted doping including incorporation of single Mn$^{2+}$ ions into the QDs. Then, 4 layer-pairs for the upper DBR are deposited, completing the growth, which increases the light-matter coupling compared to sample studied in Ref.\cite{PacuskiCGD17}{}. To further improve the in-coupling of the optical excitation and the out-coupling of the optical signals, we attach a solid immersion lens (SIL) made of zirconium oxide on the sample surface. This half-ball lens with a 500~\textmu m diameter allows to perform optical microscopy up to approximately 50~\textmu m away from its axis without introducing significant geometrical aberration. The SIL increases the numerical aperture (NA) of the beam in the semiconductor material by reducing the refraction when crossing the sample surface. It further decreases the spherical diffraction resulting from the excitation fields passing through the semiconductor-air interface; at the same time it reduces the total internal reflection of the emitted FWM signal on the way back.

Monitoring micro-photoluminescence (\textmu PL) at $T=7$~K between energies $E_{\rm X}=1.85$~eV ($\lambda_{\rm X}=670$~nm) and 1.80~eV (690~nm), we observe the recombinations of individual excitons localized at the interface fluctuations, forming weakly-confined QDs, similarly as in GaAs structures.\cite{LangbeinPRL05,KasprzakNatPhot11} In the white light reflectance in Fig.~\ref{fig:sample}(b) we identify the cavity mode centered between $E_{\rm cav}=1.85$~eV ($\lambda_{\rm cav}=670$~nm) and 1.81~eV (685~nm) depending on the investigated position on the sample with a full-width at half maximum (FWHM) of 12.5~nm, yielding the quality factor of $Q = \Delta E_{\rm cav}/E_{\rm cav}\approx 55$.

\section{Undoped quantum dot}

\begin{figure}[h]
    \centering
    \includegraphics[width=0.55\columnwidth]{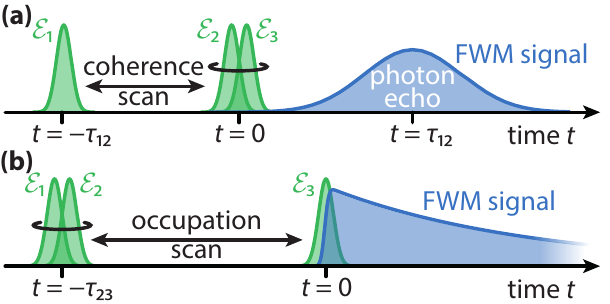}
   \caption{Schematic picture of the performed FWM experiments. (a) Coherence scan by varying $\tau_{12}$ resulting in the photon echo formation. (b) Occupation scan by varying $\tau_{23}$ exhibiting a typical exponential decay.}
    \label{fig:pulses}
\end{figure}%

\begin{figure}[h]
    \centering
    \includegraphics[width=0.65\columnwidth]{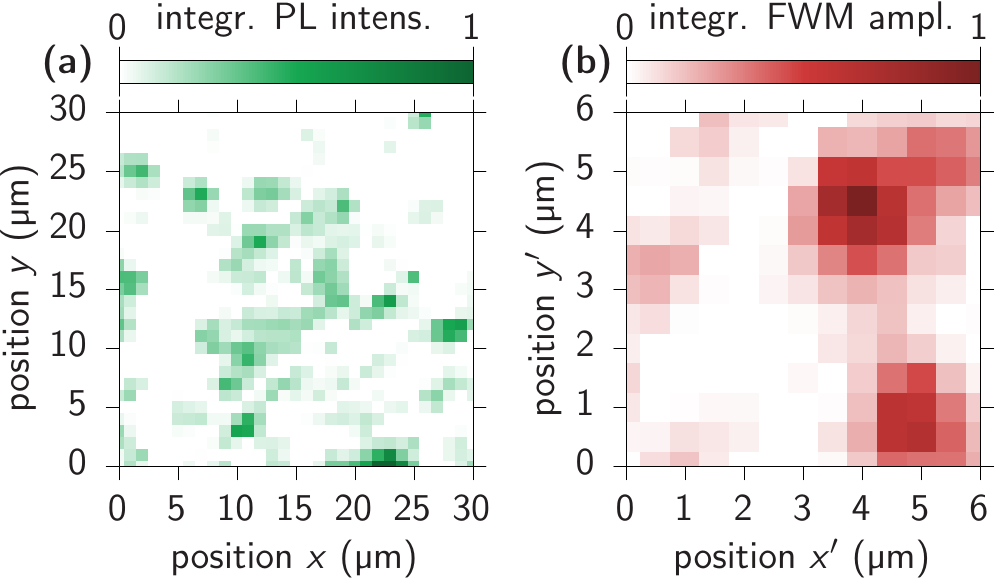}
   \caption{Spatial mapping of the PL in (a) and the FWM in (b). The scans reveal QDs with excitons that are efficiently coupling to the optical excitation. Note, that the detected areas on the sample are not aligned.}
    \label{fig:scan}
\end{figure}%

To perform FWM microscopy we use a laser pulse train centered around $\lambda=680$~nm at the repetition rate of 76~MHz, generated by an optical parametric oscillator ({\it Inspire 50} by {\it Radiantis}) pumped by a femtosecond Ti:Sapphire oscillator ({\it Tsunami-Femto} by {\it Spectra-Physics}). To induce FWM, we generate three beams $\Eo$, with respective inter-pulse delays $\tau_{12}$ and $\tau_{23}$ as schematically shown in Fig.~\ref{fig:pulses}, introduced by a pair of mechanical delay stages. The beams pass through acousto-optic modulators where they undergo distinct shifts $\Omega_{1,2,3}$ of the carrier frequency. Using a microscope objective ({\it Olympus}, NA=0.6), the beams are focused reaching a diffraction limited spot on the surface of the sample. The sample is placed in a helium-flow cryostat operating at $T=7$~K. By raster scanning the position of the objective, we can construct hyperspectral images of the optical signals.\cite{KasprzakPSSb09, FrasNatPhot16} Fig.~\ref{fig:scan} shows exemplary scans of the PL and the FWM signal in (a) and (b), respectively, where the maps were accumulated for a range of transition energies. A pulse shaper is used to correct the temporal chirp, mainly stemming from the thick optics in the acousto-optic modulators and the objective, to attain transform-limited pulses of around 150~fs duration. The reflected light from the sample is collected by the same objective and directed into an imaging spectrometer with a CCD camera at its output. The FWM response, which in the lowest (third) order is proportional to $\Ea^{\ast}\Eb\Ec$, propagates shifted by the radio-frequency $\Omega_{\rm FWM}=\Omega_3+\Omega_2-\Omega_1$, which is around 80~MHz. Its amplitude and phase are thus obtained via optical heterodyning the reflected light at $\Omega_{\rm FWM}$. Additionally, a reference beam $\Er$ is employed to perform spectral interferometry. More details about the experimental setup can be found in Refs.\cite{FrasNatPhot16, JakubczykACSNano19}{}. As explained below, by inspecting FWM temporal and delay dynamics, we obtain full information regarding the system's inhomogeneous $\sigma$ and homogeneous $\gamma=2\hbar/T_2$ dephasing, as well as its population decay.

\begin{figure}[t]
    \centering
    \includegraphics[width=0.55\columnwidth]{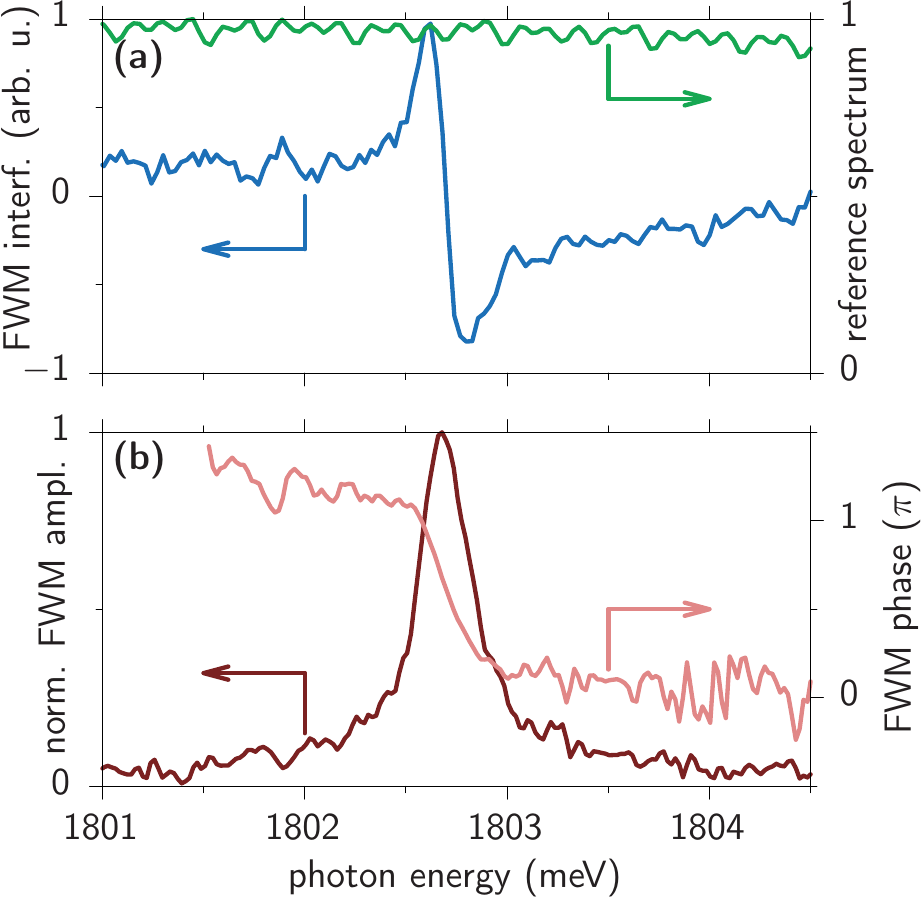}
   \caption{FWM spectroscopy of a QD exciton. (a) Spectrum of the reference pulse in green and a typical spectral heterodyne interferogram in blue. (b) FWM amplitude spectrum in dark red and the FWM phase in pale red.}
    \label{fig:X_spec}
\end{figure}

For our FWM investigations, we select those optical transitions that dominate in PL and that are spectrally located at the center of the cavity mode. In Fig.~\ref{fig:X_spec}(a) we present a typical spectral interference (blue) heterodyned at $\Omega_{\rm FWM}$ originating from a undoped CdTe QD together with the laser pulse spectrum (green). The retrieved FWM amplitude and phase are presented in Fig.~\ref{fig:X_spec}(b) as dark and pale red line, respectively. While the FWM amplitude exhibits a typical peak structure, the respective phase shows a jump of approximately $\pi$~\cite{KasprzakNatPhot11}. It is worth to note that QDs generating FWM are rather isolated with typical distances of several \textmu m, as exemplified by the FWM hyperspectral mapping presented in Fig.~\ref{fig:scan}(b).

\begin{figure}[t]
    \centering
    \includegraphics[width=0.5\columnwidth]{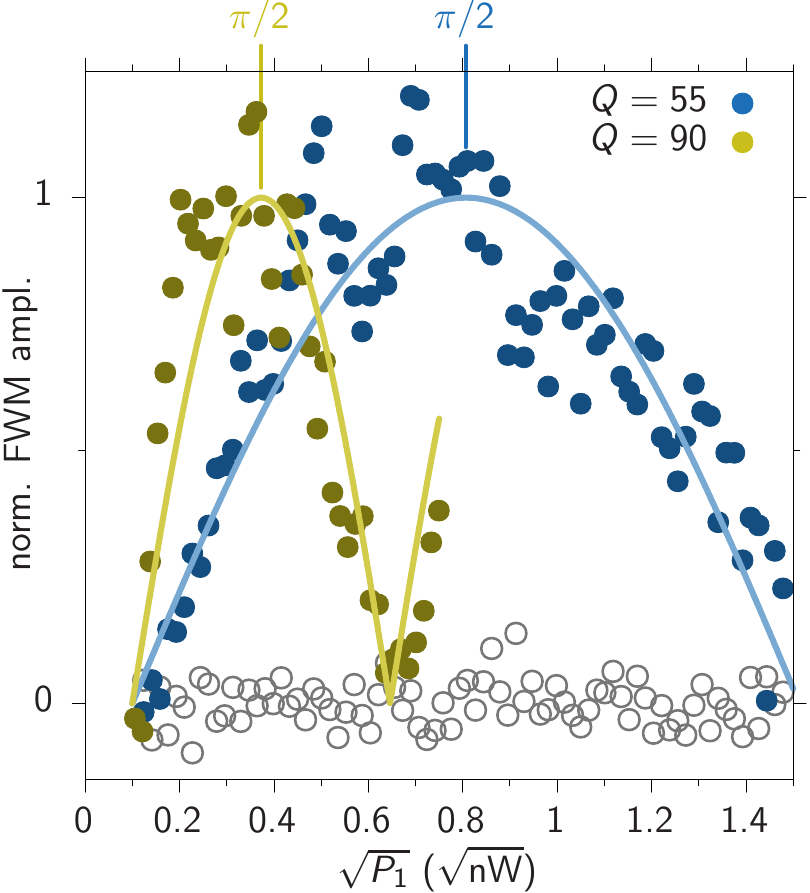}
   \caption{Rabi rotations. FWM amplitude as a function of the applied peak field amplitude of the first laser pulse $\sqrt{P_1}$ while $P_{2,3}$ are fixed. Measurement as dark dots and fits with $\left|\sin(\theta_1-\theta_0)\right|$ as pale lines for the cavity quality factors $Q=55$ (blue) and $Q=90$ (yellow).}
    \label{fig:Rabi}
\end{figure}

To first characterize the enhanced light-matter coupling, we measure how the FWM amplitude depends on the applied laser pulse intensities. We thus fix the excitation powers of $\Eb$ and $\Ec$ to $P_2=P_3=0.25$~\textmu W and vary $\Ea$'s power $P_1$. In Fig.~\ref{fig:Rabi} we plot the spectrally integrated FWM amplitude as a function of $\sqrt{P_1}$, which is proportional to the pulse area $\theta_1=\int dt\,\mathcal E_1(t)/\hbar$, where $\mathcal E_1$ is the electric field amplitude of the first pulse at the QD location multiplied by the transition dipole matrix element. As the measured FWM amplitude (blue dots) is proportional to the microscopic coherence of the exciton state,\cite{WiggerPRB17,WiggerOptica18} the signal is proportional to $\left|\sin(\theta_1-\theta_0)\right|$ (blue line), i.e., a Rabi rotation is detected with a maximum corresponding to $\theta_1-\theta_0=\pi/2$ at $P_1=0.55$~\textmu W. The offset $\theta_0$ might stem from imperfect reflection of the photonic structure and internal absorption. With respect to our previous experiments performed on a half-cavity structure\cite{PacuskiCGD17} with $Q\approx20$, a $\pi/2$ pulse area is here attained for an around 6 times weaker impinging laser power. Such an enhanced coupling between the external excitation and the QD exciton is due to the moderately larger Q-factor of the microcavity and thus a larger effective $\mathcal E_j$ for the same external $P_j$. To further demonstrate the correspondence between the cavity Q-factor and the powers required to reach a $\pi/2$ pulse, we measure the FWM's intensity dependence on a similar microcavity sample, that is fabricated from 12 (6) stacked DBR-pairs at the bottom (top). As a result its Q-factor reaches 90. From the measured Rabi flopping (yellow points and line) we deduce that the pulse area of $\pi/2$ corresponds to $P_1=0.12$~\textmu W. We note however that a further increase of $Q$ does not necessarily lead to a better light-matter coupling. This is due to spectral filtering of the incoming excitation pulses and a respective increase of their temporal duration inside the cavity.\cite{WiggerOptica18} An optimal light-matter coupling is achieved when the spectral widths of the cavity mode and of the driving pulses are matched.

\begin{figure}[h]
    \centering
    \includegraphics[width=0.65\columnwidth]{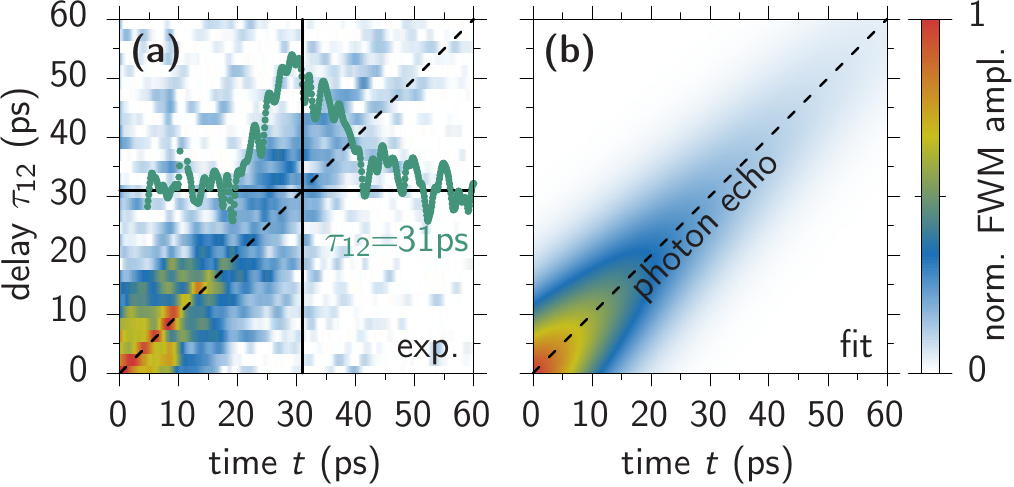}
   \caption{Photon echo formation. FWM dynamics as a function of time $t$ after the third pulse and the delay $\tau_{12}$. (a) Measurement and (b) fit with Eq.~\eqref{eq:deph}. The green points in (a) show the measurement at $\tau_{12}=31$~ps.}
    \label{fig:echo}
\end{figure}

We now shift the investigation to the temporal domain. A typical pulse sequence of the experiment is presented in Fig.~\ref{fig:pulses}, where the signal is generated after the arrival of all three pulses $\Eo$. In inhomogeneously broadened systems, the FWM signal for $\tau_{12}>0$ (see Fig.~\ref{fig:pulses}(a)) forms a photon echo.\cite{LangbeinPRL05} Even though the echo formation is commonly known for ensembles of emitters, it can also be generated for individual transitions. Here, the photon echo arises due to the exciton's stochastic spectral wandering in time, accumulating into an effective inhomogeneous broadening of width $\sigma$ in the time-averaged heterodyne experiment.\cite{PattonPRB06, KasprzakNJP13, MermillodPRL16, HahnAdvSci21} To illustrate that, in Fig.~\ref{fig:echo}(a) we show the measured time-resolved FWM amplitude as a function of the time after the third pulse $t$ and the delay $\tau_{12}$, while fixing $\tau_{23}=0$. We observe that with increasing delay $\tau_{12}$ the maximum of the signal shifts in time $t$ along the diagonal $\tau_{12}=t$ (dashed line). We see that for $\tau_{12}>20$~ps, the echo is fully developed, i.e., the FWM signal takes the form of a Gaussian transient. This is exemplarily shown by the time-resolved FWM amplitude measured at $\tau_{12}=31$~ps (green dots). By fitting the entire FWM dynamics with
\begin{align}
	S(t,\tau_{12}) \sim \exp\left[-\frac{t+\tau_{12}}{T_2}\right] \exp\left[-\frac{(t-\tau_{12})^2}{2T_\sigma^2}\right] \label{eq:deph}
\end{align}%
as shown in Fig.~\ref{fig:echo}(b) we directly retrieve the homogeneous dephasing time $T_2=(36.5\pm0.2)$~ps and the inhomogeneous dephasing time $T_\sigma=(9.25\pm0.05)$ ps. These times directly correspond to spectral broadenings of $\gamma = 2\hbar/T_2 \approx (36.1\pm0.2)$~\textmu eV and $\sigma = 2\hbar/T_\sigma \approx (142\pm1)$~\textmu eV.\cite{KasprzakNJP13, Jakubczyk2DMat18}

\begin{figure}[h]
    \centering
    \includegraphics[width=0.55\columnwidth]{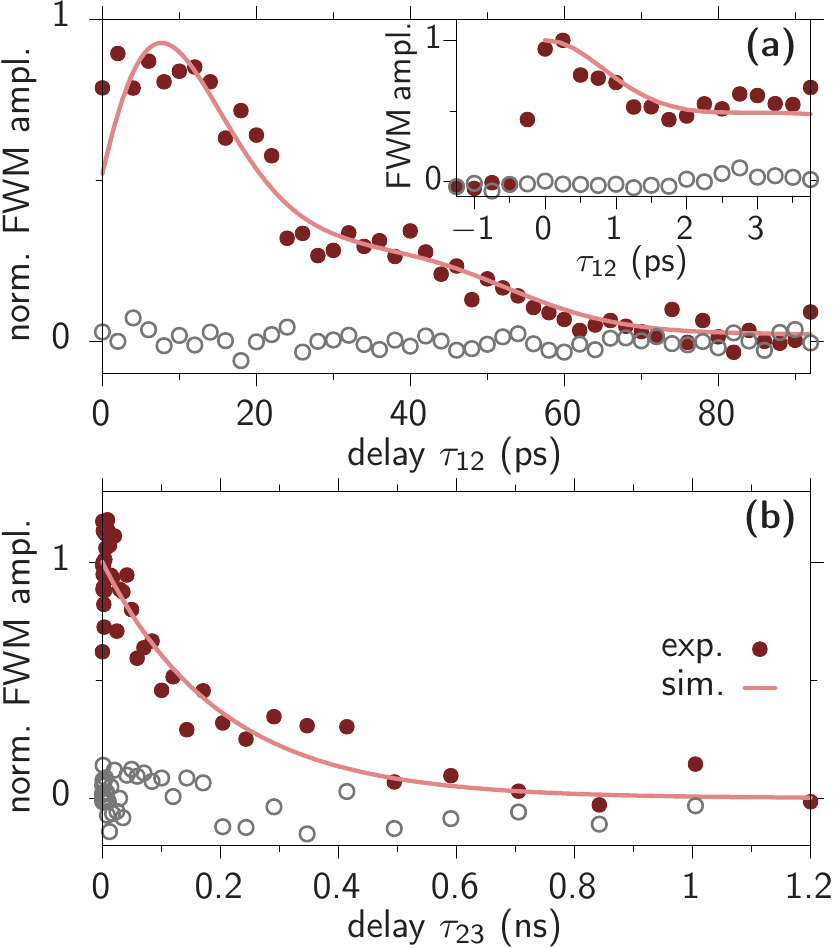}
   \caption{FWM dynamics of the QD exciton. (a) Coherence dynamics as a function of the delay $\tau_{12}$ exhibiting dephasing and FSS-induced beats. The inset shows the PID on a few ps time scale. (b) Population dynamics as a function of the delay $\tau_{23}$ exhibiting a single exponential decay.}
    \label{fig:X_dyn}
\end{figure}

To have a closer look at the coherence dynamics of the exciton, we measure the time-integrated FWM signal as a function of $\tau_{12}$, depicted in Fig.~\ref{fig:X_dyn}~(a) as dark red dots. The signal shows the expected behavior after time integrating the photon echo in Eq.~\eqref{eq:deph} (Fig.~\ref{fig:echo}) over $t$, which consists of an exponential decay that dominates for large delays ($\tau_{12}\gg T_\sigma$) and an increasing contribution during the development of the full echo for $\tau_{12}<T_\sigma$. In addition we find a modulation of the signal stemming from the FFS of the two linearly polarized excitons in the QD. As the linearly polarized $\Eo$ are misaligned from the anisotropy axes of the QD, both excitons are excited and the corresponding coherences contribute to the final FWM signal.\cite{PattonPRB06, KasprzakNJP13, MermillodOptica16} With the model described  in Ref.~\cite{MermillodOptica16} we can fit the measured data and retrieve the pale red line with a FFS of $\delta_{\rm FFS} = \hbar 2 \pi/T_\delta = (82\pm 5)$~\textmu eV and a light polarization angle of $\alpha = (25\pm 1)^\circ$ with respect to one of the QD excitons. An exemplary FWM spectrum exhibiting a large FFS can be found the Supporting Information Fig. S1. We note that the exciton-biexciton transition is not covered by $\Eo$ and therefore does not influence the dynamics.\cite{MermillodOptica16} Examples of neutral exciton-biexciton complexes with both bound and unbound character, typical for weakly-confined QDs,\cite{KasprzakJOSAB12} are readily identified in FWM on the same sample, as shown for a bound example in the Supporting Information Fig. S2.

In the inset of Fig.~\ref{fig:X_dyn}(a) the FWM dynamics are shown for a delay timescale of a few picoseconds. After the signal's rise from negative delays, it reaches a maximum around $\tau_{12}=0$. After that it drops within less than 2~ps to approximately 0.5 of its maximum value. This fast decay is recognized as PID, due to the optical excitation with pulses that are siginificantly shorter than the polaron formation process.\cite{JakubczykACSPhot16, WiggerOL20} The FWM dynamics are reproduced by the depicted simulation (pale red line) in the well established independent boson model.\cite{mahan, JakubczykACSPhot16} In this model the exciton-phonons coupling is described by additional dynamics of the exciton coherence, in the form of the PID function $\tilde{p}_{\rm PID}$. The full FWM dynamics are therefore given by
\begin{align}
	p_{\rm FWM}(t,\tau_{12}) \sim \tilde{p}_{\rm PID}(t,\tau_{12}) S(t,\tau_{12})\,,
\end{align}%
where $S(t,\tau_{12})$ is the homogeneous and inhomogeneous dephasing contribution from Eq.~\eqref{eq:deph}. For optical pulses that are much shorter than the considered phonon periods, the PID dynamics can be calculated analytically in the limit of ultrafast pulses via\cite{VagovPRB02}
\begin{align}
	&\tilde{p}_{\rm PID}(t,\tau_{12}) \notag\\
	&= \exp\bigg\{\left|\frac{g_{\bm q}}{\omega_{\bm q}}\right|^2 \Big[2\cos(\omega_{\bm q}t)-3 + e^{i\omega_{\bm q} \tau_{12}}(2-e^{i\omega_{\bm q}\tau_{12}})\notag\\
		&\qquad - N_{\bf q}\left|e^{i\omega_{\bm q}\tau_{12}}(2-e^{i\omega_{\bm q}t})-1\right|^2\Big]\bigg\}\,,
\end{align}%
with the thermal occupation of the phonon modes $N_{\bm q} = \{\exp[\hbar\omega_{\bm q}/(k_BT)]-1\}^{-1}$. For simplicity we here assume a spherical exciton wave function for which the coupling constant can be written as 
\begin{align}
	g_{\bf q} = \frac{q D}{\sqrt{2\rho\hbar V\omega_{\bm q}}}e^{-\frac 12 q^2 a^2}\,,
\end{align}%
with the normalization volume $V$. For the material parameters we use the mass density of $\rho=5870$kg/m$^3$, an effective deformation potential strength of $D=9$~eV,\cite{KranzerJPD73} and assume an isotropic phonon dispersion $\omega_{\bm q}=c q$ with the longitudinal acoustic sound velocity $c=3.2$~nm/ps.\cite{KrischPRB97} We find the best agreement with the measured FWM dynamics for an exciton localization length of $a=2$~nm.

To complete the study of the undoped QD, in Fig.~\ref{fig:X_dyn}(b) we present the exciton occupation dynamics, which are measured by the $\tau_{23}$-dependence of the FWM amplitude while fixing $\tau_{12}=0$ (red dots). An exponential decay (pale red line) is observed with a decay time of $T_1=(200\pm 25)$~ps. This decay is attributed to the radiative recombination of the bright exciton states. The decay of the dark exciton typically happens on a much longer timescale of a few tens of ns\cite{SmolenskiPRB12} and is therefore not resolved here. The collection of the QD parameters ascertained by the FWM study is gathered in Table~\ref{tab:table1}.

\begin{table}[h]
    \caption{Parameters characterizing the optical  properties of the QD exciton retrieved by FWM spectroscopy.}
    \label{tab:table1}
    \centering
    \begin{tabular}{l | l}
    \hline
    homogeneous broadening   &  $\gamma=(36.1\pm 0.2)$~\textmu eV \\
    inhomogeneous broadening & $\sigma=(142\pm 1)$~\textmu eV\\
    bright exciton lifetime &  $T_1=(200\pm25)$~ps \\
    fine-structure splitting  &  $\delta_{\rm FSS}=(82\pm 5)$~\textmu eV \\
    light-matter coupling &  $\theta=\pi/2$ @ $P=0.55$~\textmu W \\
    \hline
    \end{tabular}
\end{table}

\section{Mn-doped quantum dot}

\begin{figure}[t]
    \centering
    \includegraphics[width=0.55\columnwidth]{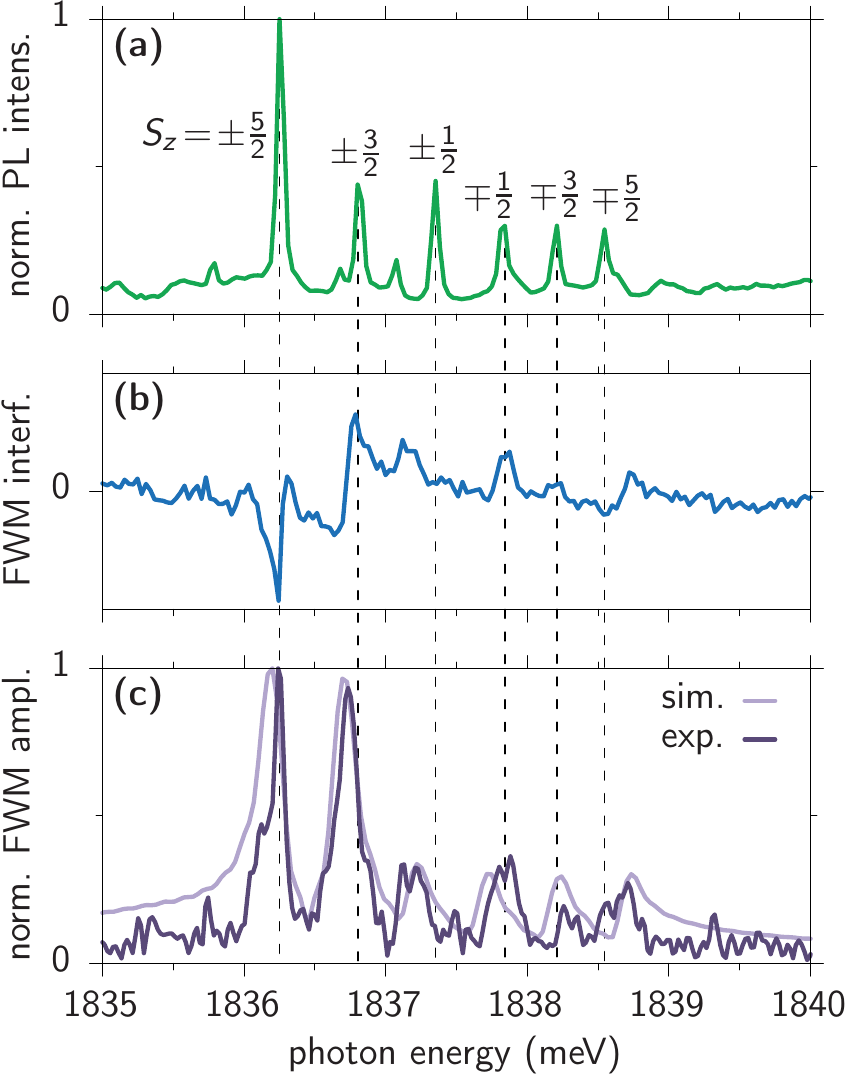}
   \caption{Spectral characterization of a Mn-doped QD. (a) PL spectrum exhibiting the six spectral lines induced by the Mn dopant. (b) FWM spectral interferogram. (c) FWM amplitude spectrum with the measurement in dark and the simulation in pale violet.}
    \label{fig:Mn_spec}
\end{figure}

After this characterization of the excitonic properties, we come to the QD containing a single Mn$^{2+}$ ion. Such a QD is recognized by measuring a PL spectrum as presented in Fig.~\ref{fig:Mn_spec}(a). The insertion of an individual Mn$^{2+}$ ion into a QD, within the volume of the exciton's wave function, is confirmed by detecting the comb of six separate spectral lines,\cite{BesombesPRL04, GorycaPRL09} as shown in the PL spectrum in Fig.~\ref{fig:Mn_spec}(a), which is characteristic for a sufficiently symmetric QD when the exciton-Mn exchange interaction dominates over the  anisotropic electron-hole exchange interaction, i.e., when the splitting of the lines due to the exciton-Mn interaction is larger than the fine-structure splitting.\cite{LegerPRL05} The exciton transition is sensitive to the spin state of the magnetic ion: The exchange interaction between the QD exciton and the ion leads to spin-dependent spectral shifts with respect to the undoped QD exciton in the range of a few meV. The electron-Mn exchange interaction furthermore leads to spin flips resulting in a coupling between bright and dark excitons which, however, typically becomes effective only at high magnetic fields.\cite{BesombesPRL04, FernandezPRB06} Without an additional magnetic field, the Mn spin projection $S_z$ freely jumps between its possible realizations, namely $\pm\frac{5}{2}$, $\pm\frac{3}{2}$, and $\pm\frac{1}{2}$. In a time averaged measurement, this results in the development of six spectral components. 

\begin{figure}[t]
    \centering
    \includegraphics[width=0.55\columnwidth]{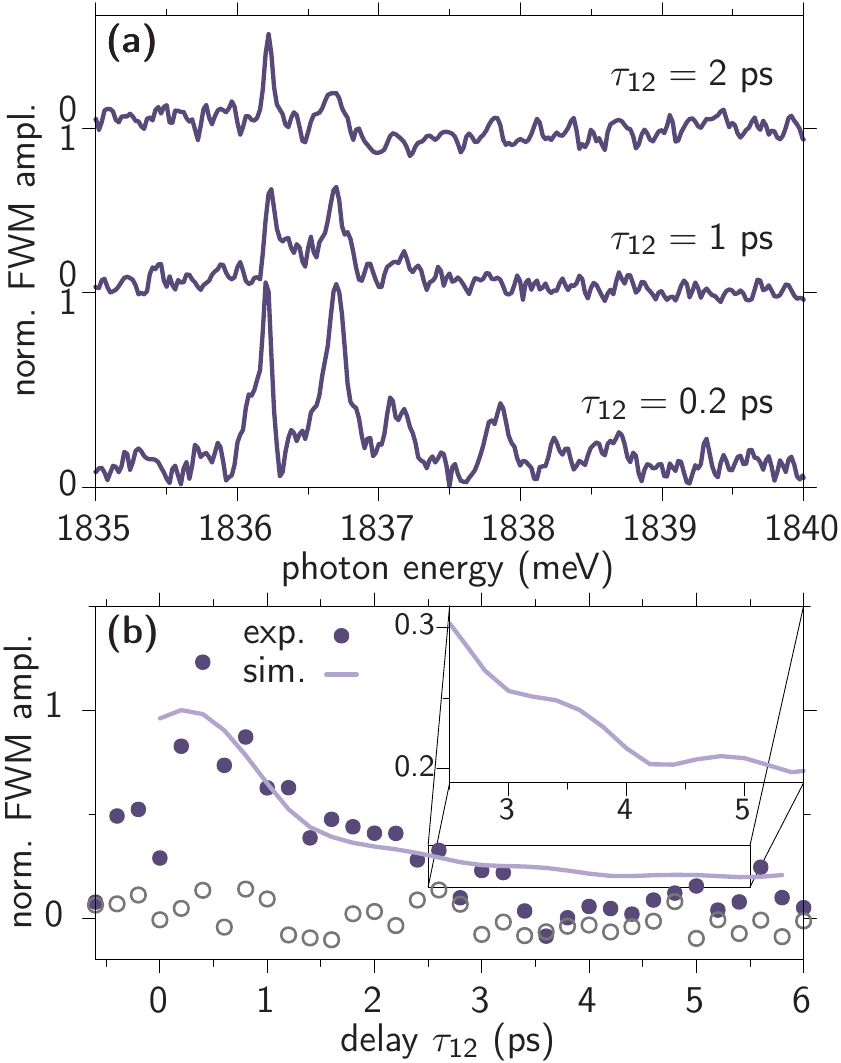}
   \caption{Coherence dynamics of a Mn-doped QD. (a) Evolution of the FWM spectrum with increasing delay $\tau_{12}$ from bottom to top. (b) Integrated FWM amplitude as a function of the delay $\tau_{12}$ with the experiment as dark violet dots and the simulation as pale violet line.}
    \label{fig:Mn_dyn}
\end{figure}

The FWM spectral interferogram and the resulting FWM amplitude are shown in Fig.~\ref{fig:Mn_spec}(b) and (c), respectively. Here, we also recover six spectral lines with their amplitudes increasing for smaller transition energies. Interestingly, fluctuations of such a single spin generate a peculiar type of inhomogenous broadening acting on the exciton. A typical Mn spin-flip time in a CdTe QD is on the order of several \textmu s,\cite{BesombesPRL04, GorycaPRL09, GorycaPRB15} which is much longer than the measured exciton lifetime of 200~ps. During the integration time of 10~ms, the exciton performs a few thousand spectral jumps. Because the spin-flip time is much longer than the exciton lifetime, the corresponding random jumps of the transitions energy can be interpreted as a discrete ensemble. In Fig.~\ref{fig:Mn_dyn} we examine how the FWM signal behaves as a function of the delay $\tau_{12}$. In Fig.~\ref{fig:Mn_dyn}(a) we show examples of the measured FWM spectra for three different delays $\tau_{12}$ increasing from bottom to top as labeled in the plot. We already see that the overall amplitude of the signal strongly decreases with larger delays. In Fig.~\ref{fig:Mn_dyn}(b) we show the integrated FWM amplitude as a function of $\tau_{12}$ as violet dots, which confirms the rapid drop of the system's coherence. The significant decoherence within the first 2~ps is already known from the undoped QD and it stems from the PID in Fig.~\ref{fig:X_dyn}(b, inset). After the PID, i.e., for $\tau_{12} > 1.5$~ps only long-time dephasings (homogeneous and inhomogeneous) reduce the signal. In the simulation depicted as pale violet line, we take the impact of the Mn ion into account by calculating an ensemble (ens) average of the six transition energies $\hbar\Delta\omega_{n}$ via
\begin{align}
	p_{\rm FWM}^{\rm ens}(t,\tau_{12}) = \sum_{n=1}^{6} p^{(n)}_{\rm FWM}(t,\tau_{12})e^{-i\Delta \omega_{n} (t-\tau_{12})}\,,
\end{align}%
where each FWM contribution $p^{(n)}_{\rm FWM}(t,\tau_{12})$ can be individually weighted. These weights of the six contributions are chosen such that the simulated spectral distribution (pale violet line) in Fig.~\ref{fig:Mn_spec}(c) agrees with the measured one (dark line). In the resulting coherence dynamics in Fig.~\ref{fig:Mn_dyn}(b) the discrete equidistant ensemble results in a slight beating of the FWM signal. This oscillation is highlighted by the inset, which is a zoom-in on the black rectangle. The effect is similar to the FSS beat observed from the undoped dot in Fig.~\ref{fig:X_dyn}(a). However, here it stems from all possible frequency differences in the six-state ensemble.

\section{Conclusions}
In this work, we have studied the coherence properties of an exciton confined to a CdTe QD by FWM spectroscopy. The creation and detection of this nonlinear optical signal, scaling with the third power of the investigated dipole moment of the quantum system, required the incorporation into a low-Q DBR cavity. To finally reach the required efficiency for the light-matter coupling we additionally applied a solid immersion lens on the sample surface. With this setup we were able to detect photon echo dynamics, which allowed us to determine the homogeneous and inhomogeneous dephasing of the QD exciton. We further measured beats of the signal, revealing the fine-structure splitting of the linearly polarized excitons, and their population lifetime. Considering a QD hosting a single Mn-dopant we performed the first FWM spectroscopy study of the characteristic six-lined spectral structure. We showed that this unique shape of transition energies, stemming from random fluctuations between the Mn-spin states, results in additional dephasing dynamics and a signal beating in the ensemble average.

As the Mn spin-orientation can be controlled by an external magnetic field, forthcoming magneto-FWM micro-spectroscopy experiments will allow to further monitor the impact of the observed discrete inhomogeneous broadening onto the exciton coherence dynamics. It will furthermore allow to study the coherence dynamics associated with the exchange-induced coupling between bright and dark excitons giving rise to the characteristic anticrossings in the magneto-PL of Mn-doped QDs.\cite{BesombesPRL04} This progress will reveal the system's potential for a coherent ultrafast spin-photon interface. In particular 2D FWM spectroscopy will unveil internal coherent interactions in the coupled exciton-Mn$^{2+}$ system. It is possible to shift the Mn-doped QD emission energies below 1770~meV (wavelengths above 700~nm), see PL spectrum in Fig. S3 of the Supporting Information. Further work regarding the growth will optimize the sample performance at this spectral range, enabling implementation of the resonant spectroscopy employing standard Ti:Sapphire femtosecond laser sources. At this point, we can combine more sophisticated and innovative photonic nanostructures, currently emerging for the CdTe-platform,\cite{BoguckiLight20} with QDs that can already contain a variety of different magnetic impurities.\cite{KobakNatComm14, SmolenskiNatComm16} These technological and spectroscopic advances open exciting prospects for coherent nonlinear spectroscopy of hybrid exciton-spin systems in semiconductor nanostructures.

\begin{acknowledgement}

This work was partially supported by the Polish National Science Centre (NCN) under decision DEC-2015/18/E/ST3/00559. 
Tomasz Jakubczyk acknowledges support from the Polish National Agency for Academic Exchange (NAWA) under Polish Returns 2019 programme (Grant No. PPN/PPO/2019/1/00045/U/0001). Daniel Wigger thanks NAWA for financial support within the ULAM program (Grant No. PPN/ULM/2019/1/00064).

\end{acknowledgement}

\begin{suppinfo}

Supporting Information contains:\\
-- Exemplary FWM spectra showing individual excitons with a large fine-structure splitting.\\
-- Two-dimensional FWM spectrum showing the exciton-biexciton complex.\\
-- Photoluminescence spectrum of a Mn-doped QD emitting around 1710~meV.

\end{suppinfo}


\providecommand{\latin}[1]{#1}
\makeatletter
\providecommand{\doi}
  {\begingroup\let\do\@makeother\dospecials
  \catcode`\{=1 \catcode`\}=2 \doi@aux}
\providecommand{\doi@aux}[1]{\endgroup\texttt{#1}}
\makeatother
\providecommand*\mcitethebibliography{\thebibliography}
\csname @ifundefined\endcsname{endmcitethebibliography}
  {\let\endmcitethebibliography\endthebibliography}{}

\end{document}





\thispagestyle{empty}
\setcounter{page}{0}
\pagebreak
~\\[-10mm]
\begin{figure}[t]
    \centering
    \includegraphics[width=0.5\columnwidth]{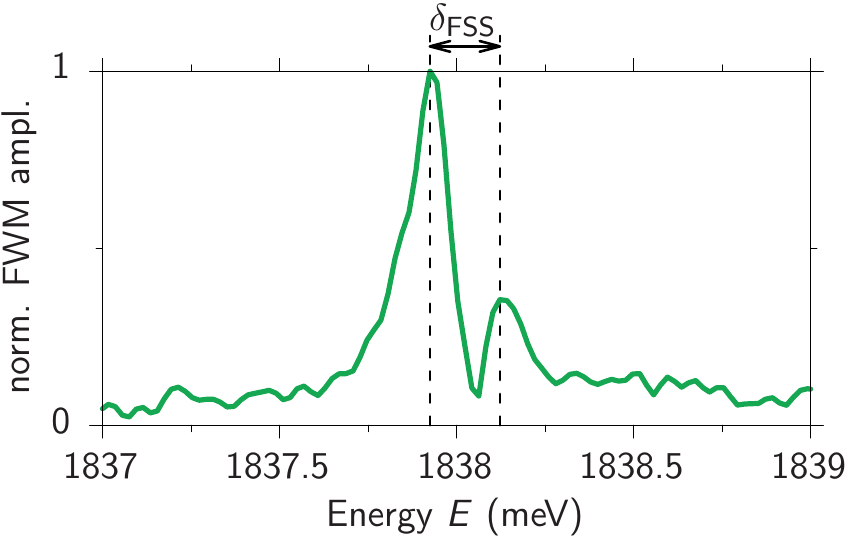}
   \caption{Four-wave mixing spectrum of an un-doped quantum dot showing a fine-structure splitting of $\delta_{\rm FSS} \approx 0.2$~meV.}
    \label{fig:FSS}
\end{figure}%
~\\[-10mm]
\begin{figure}[h]
    \centering
    \includegraphics[width=0.75\columnwidth]{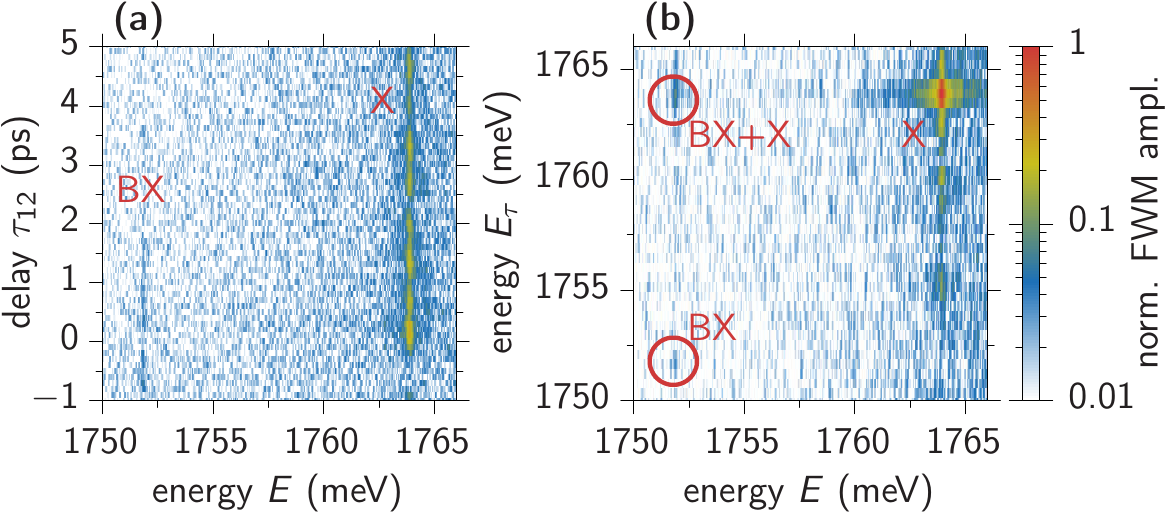}
   \caption{(a) Delay scan of an un-doped quantum dot showing an exciton (X) biexciton (BX) complex. The BX clearly appears for $\tau_{12}<0$. (b) Corresponding 2D spectrum demonstrating the coherent coupling between X and BX by the off-diagonal spectral peak. The biexciton binding energy (BBE) is $\delta_{\rm BBE}\approx12$~meV.}
    \label{fig:2D}
\end{figure}%
~\\[-10mm]
\begin{figure}[h!]
    \centering
    \includegraphics[width=0.5\columnwidth]{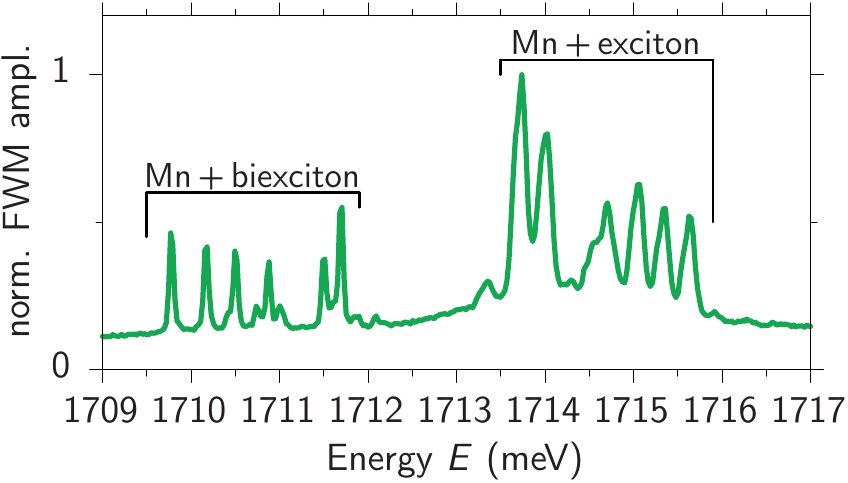}
   \caption{Photoluminescence spectrum of a Mn-doped quantum dot showing the exciton and biexciton manifolds at around 1715~meV and 1710~meV, respectively.}
    \label{fig:2D}
\end{figure}%

\bibliography{Mn}